\documentclass[useAMS,usenatbib]{mn2e}
\usepackage{rotating}
\usepackage{subfigure}
\RequirePackage{setspace}
\usepackage{graphicx}
\usepackage{epsf}
\usepackage{fancyhdr}
\usepackage{epsfig}
\usepackage{natbib}
\usepackage{amsmath}
\usepackage{amssymb}
\usepackage{aas_macros}
\usepackage{epstopdf}

\title[Limits on dust emission from z$\sim$5 LBGs ]{Limits on dust emission from z$\sim$5 LBGs and their local environments}
\author[L. J. M. Davies et. al.]{L. J. M. Davies$^{1,2}$\thanks{E-mail:
Luke.Davies@bristol.ac.uk}, M. N. Bremer$^{1}$, E. R. Stanway$^{1,3}$, E. Mannering$^{1}$, \newauthor M. D. Lehnert$^{4}$, A. Omont$^{5}$\\
$^{1}$Department of Physics, University of Bristol, H.H. Wills Physics Laboratory, Tyndall Avenue, Bristol, BS8 1TL, UK\\
$^{2}$Institute of Cosmology and Gravitation, University of Portsmouth, Dennis Sciama Building, Burnaby Road, Portsmouth PO1 3FX\\
$^{3}$Department of Physics, University of Warwick, Gibbet Hill Road, Coventry, CV4 7AL, UK\\
$^{4}$Laboratoire d'Etudes des Galaxies, Etoiles, Physique et Instrumentation GEPI, Observatoire de Paris, UMR8111 du CNRS, \\ Meudon, 92195 France\\
$^{5}$Institut dÕAstrophysique de Paris, CNRS and Universite Pierre et Marie Curie, 98bis Boulevard Arago, F-75014 Paris, France}

\begin{document}

\date{Accepted: May 2012}

\pagerange{\pageref{firstpage}--\pageref{lastpage}} \pubyear{2012}

\maketitle

\begin{abstract}

  We present 1.2\,mm MAMBO-2 observations of a field which is over-dense
  in Lyman Break Galaxies (LBGs) at $z\sim5$. The field includes seven
  spectroscopically-confirmed LBGs contained within a narrow
  ($z=4.95\pm0.08$) redshift range and an eighth at $z=5.2$. We do not
  detect any individual source to a limit of 1.6\,mJy/beam
  (2\,$\times$ rms). When stacking the flux from the positions of all
  eight galaxies, we obtain a limit to the average 1.2 mm flux of
  these sources of 0.6\,mJy/beam. This limit is consistent with FIR
  imaging in other fields which are over-dense in UV-bright galaxies
  at $z\sim5$. Independently and combined, these limits constrain the
  FIR luminosity (8-1000$\mu$m) to a typical $z\sim5$ LBG of
  L$_{\mathrm{FIR}}\lesssim 3 \times 10^{11}$\,L$_{\odot}$, implying a dust mass
  of M$_{\mathrm{dust}}\,\lesssim\,10^{8}$\,M$_{\odot}$ (both assuming a
  grey body at 30K). This L$_{\mathrm{FIR}}$ limit is an order of magnitude
  fainter than the L$_{\mathrm{FIR}}$ of lower redshift sub-mm sources
  ($z\sim1-3$). We see no emission from any other sources within the
  field at the above level. While this is not unexpected given
  millimetre source counts, the clustered LBGs trace significantly
  over-dense large scale structure in the field at $z=4.95$. The lack
  of any such detection in either this or the previous work, implies
  that massive, obscured star-forming galaxies may not always trace
  the same structures as over-densities of LBGs, at least on the
  length scale probed here. We briefly discuss the implications of
  these results for future observations with ALMA.

\end{abstract}

\begin{keywords}
galaxies: high-redshift - galaxies: starburst - galaxies: star-formation - radio continuum: galaxies
\end{keywords}

\section{Introduction}
\label{Into}

Lyman Break Galaxies (LBGs) form a substantial fraction of the
observed $z\sim5$ galaxy population
\citep[e.g][]{vanzella09,Douglas09,Douglas10}, with a comoving number
density of $\phi^* \sim 10^{-3}\,h^3$ Mpc$^{-3}$ \citep[at
  $z\sim4-6$, ][]{Bouwens07}. They are identified via their bright UV
continuum emission arising from hot, young stars in unobscured
starburst regions. Considerable work has been carried out to
investigate LBG properties which can be discerned from their
rest-frame UV spectra and UV/optical spectral energy distributions
\citep[e.g][]{Verma07,Stark09}.  However, until now little work has
been carried out to explore their cooler dust and interstellar gas
components necessary for a clearer and fuller picture of these
important probes of early galaxy evolution. In order to fully
understand star-formation activity at high redshift and the subsequent
evolution of early star-forming galaxies, we
need to observe their complete baryonic budget. By comparing the
stellar, molecular gas and dust fractions of these
systems we can infer their star-formation history and potential fate, therefore understanding the importance of this
population to galaxy evolution in general.

Until recently, studies of the dust content of the highest redshift
galaxies \textbf{($z\sim5$)} have been limited to massive and rare
systems in comparison to more typical and numerous star-forming
galaxies at $z\sim5$.  These studies have primarily targeted quasar
host galaxies \citep[e.g][]{Wang08,Carilli10} and the highest redshift
luminous submillimetre galaxies
\citep[SMGs, e.g][]{Coppin10,Riechers10}, a small subset of the SMG
population which is dominated by galaxies at $1<z<3$
\citep{Blain02,Chapman05, Smail02}. Dust emission has been observed in
more typical galaxies at significant redshifts ($z\sim 1-3$) but
usually only in lensed objects
\citep[e.g][]{Baker01,Negrello10,Conley11}, or from rare LBGs with
exceptionally high star-formation rates \citep[][]{Chapman09}. Given
their comparatively high source density, $z\sim 5$ LBGs are likely to
be better tracers of more typical high redshift star-formation than
such rare and highly luminous sources, even given the star-formation
rates of 10-100 M$_\odot$yr$^{-1}$ determined from the UV emission of
$\sim$L$^*$ LBGs \citep[$e.g.$][]{Verma07}.

Given that these star-formation rates are typically an order of magnitude
less than those of high redshift SMGs, and that the surface
densities of SMGs are lower than those of $z\sim5$ LBGs, unless there is substantial
obscured star-formation associated with typical LBGs, they will not
be detected at the levels usually reached by submillimeter surveys
(several mJy at 850$\mu$m-1.2mm). To detect or constrain dust emission
from $z\sim 5$ LBGs, the average source needs to be probed to sub-mJy
levels. In addition, a large-enough number of these sources need to be sampled
to search for any rare LBG that has significant sub-mm flux. Until
recently there were no published predictions of the expected sub-mm
flux of distant LBGs based on any self-consistent modelling of the
star-formation within the population. Recently \citet{Gonzalez11}
derived predicted submm fluxes for high redshift LBGs based upon the
GALFORM semi-analytic model \citep[see ][]{Cole2000}. These predicted
fluxes are approaching the observational limits of \citet{Stanway10}
(see later), indicating that such observations are starting to
challenge theoretical predictions for early galaxy evolution.

While studies of typical unlensed star-forming galaxies at lower redshifts
($z\lesssim3$ LBGs) have yet to yield an individual detection in the
FIR, we cannot necessarily extrapolate these results to infer the
properties of the $z\sim 5$ population. Many of the properties of the
$z\sim 3$ sources are different from those of typical LBGs at $z\sim
5$ \citep{Verma07}.  The higher redshift LBGs are also likely to be the
progenitors of different current-day galaxy populations than the
$z\lesssim 3$ LBGs.  While the bulk of stars in present-day galaxies
are formed at $z<3$ \citep[$e.g.$][]{Sobral12}, ellipticals found in
the cores of galaxy clusters appear to rapidly form a significant
fraction of their stars at $z>3$
\citep[$e.g.$][]{Thomas10}. Consequently, the mm/submm characteristics
of the $z\sim 5$ LBGs could differ from those of $z\lesssim3$ LBGs,
requiring that they are directly measured rather than extrapolated
from the existing work at lower redshifts \citep[$e.g.$][]{Webb03}.

In order to further explore the $z\sim 5$ LBG population in the
mm-submm regime, we present the results of 1.2mm MAMBO-2
imaging of a field selected for a high density of
spectroscopically-confirmed LBGs which trace an apparently highly
over-dense structure in the young Universe. The field is drawn from
the ESO Remote Galaxy Survey \citep[ERGS,
][]{Douglas07,Douglas09,Douglas10}, which identified a sample of 70
spectroscopically confirmed $z\sim5$ LBGs over 10 widely spaced $\sim
40$ arcmin$^{2}$ fields. 

Two of these fields display highly significant 3-dimensional
over-densities of spectroscopically-confirmed UV-bright sources over
narrow redshift ranges ($\Delta z\sim 0.1$). The two structures occupy
no more than 4 per cent of the effective survey volume, while
containing $\sim 30$ per cent of the spectroscopically-confirmed LBGs.  Hence, the LBGs in these fields are likely to trace
significant large scale structures in an early stage of their
evolution, and may themselves contain other sub-mm luminous galaxies \citep[$e.g.$][]{Capak11}. 

The redshift and spatial clustering of sources in these
fields fortuitously make them ideal testbeds for simultaneous
observations of many LBGs and additionally probe significant early
mass over-densities.  The first of these fields was studied by
\cite{Stanway10} at 870$\mu$m using the Large Apex BOlometer CAmera
(LABOCA)   resulting in a limit of $S_{\rm 870\mu m}
<0.85$mJy for the average emission from nine $z\sim 5$ LBGs. The
MAMBO-2 observations presented here are of the second
highly-over-dense field in ERGS which cover a further eight $z=5$ LBGs
and the over-dense structure that seven of these inhabit. By doubling
the sample size of spectroscopically-confirmed LBGs targeted for their
mm/sub-mm dust emission, using a field spatially unrelated from the
first, we aim to both confirm the limit on the emission from a typical
$z\sim 5$ LBG and to search for or limit the number of potential
mm/sub-mm luminous outliers in the population.

These observations also allow the exploration of the nature of $z\sim 5$
LBGs beyond what is possible with the standard optical/near-IR
analysis \citep[e.g.][]{Douglas10}. The galaxies identified through
the Lyman Break technique are selected for their rest-frame
(presumably unobscured) UV emission which itself implies a minimum
star-formation rate (and therefore, via a set of assumptions, a
minimum FIR luminosity). However, it is entirely possible that some of
these either contain obscured stellar populations or have neighbours/companions that have similarly obscured stellar populations (which
may be more significant that the UV-emitting population). These
observations directly target the source of any obscuration independent
of the UV emission.

Throughout this paper all optical magnitudes are quoted in the AB
system \citep{Oke83}, and the cosmology used is
100\textit{h}/\textit{H}$_{0}$\,=\,70kms$^{-1}$ Mpc$^{-1}$,
$\Omega_{\Lambda}$\,=\,0.7 and $\Omega_{M}$\,=\,0.3.

\begin{figure*}
\begin{center}

\includegraphics[scale=0.29]{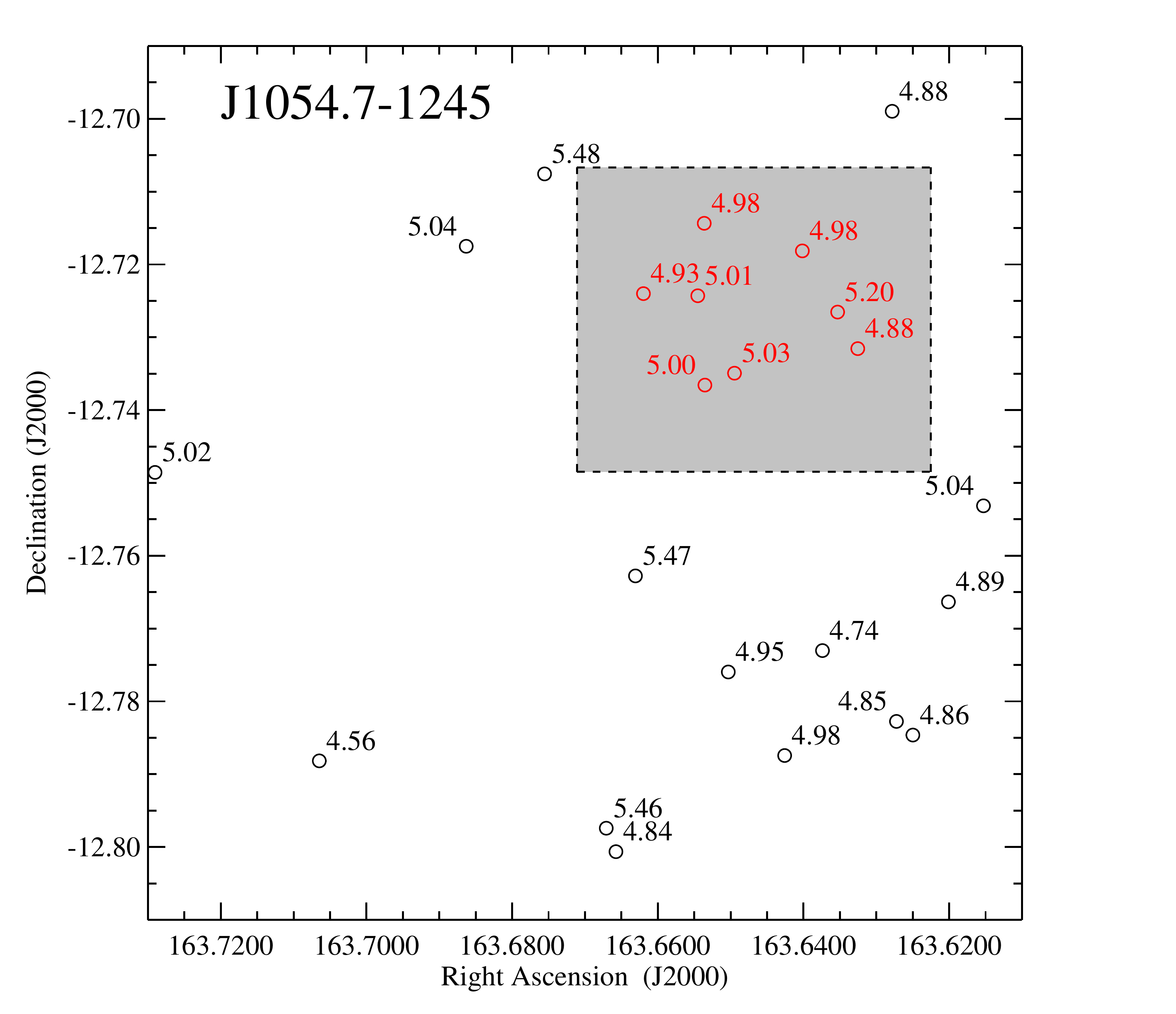}
\includegraphics[scale=0.29]{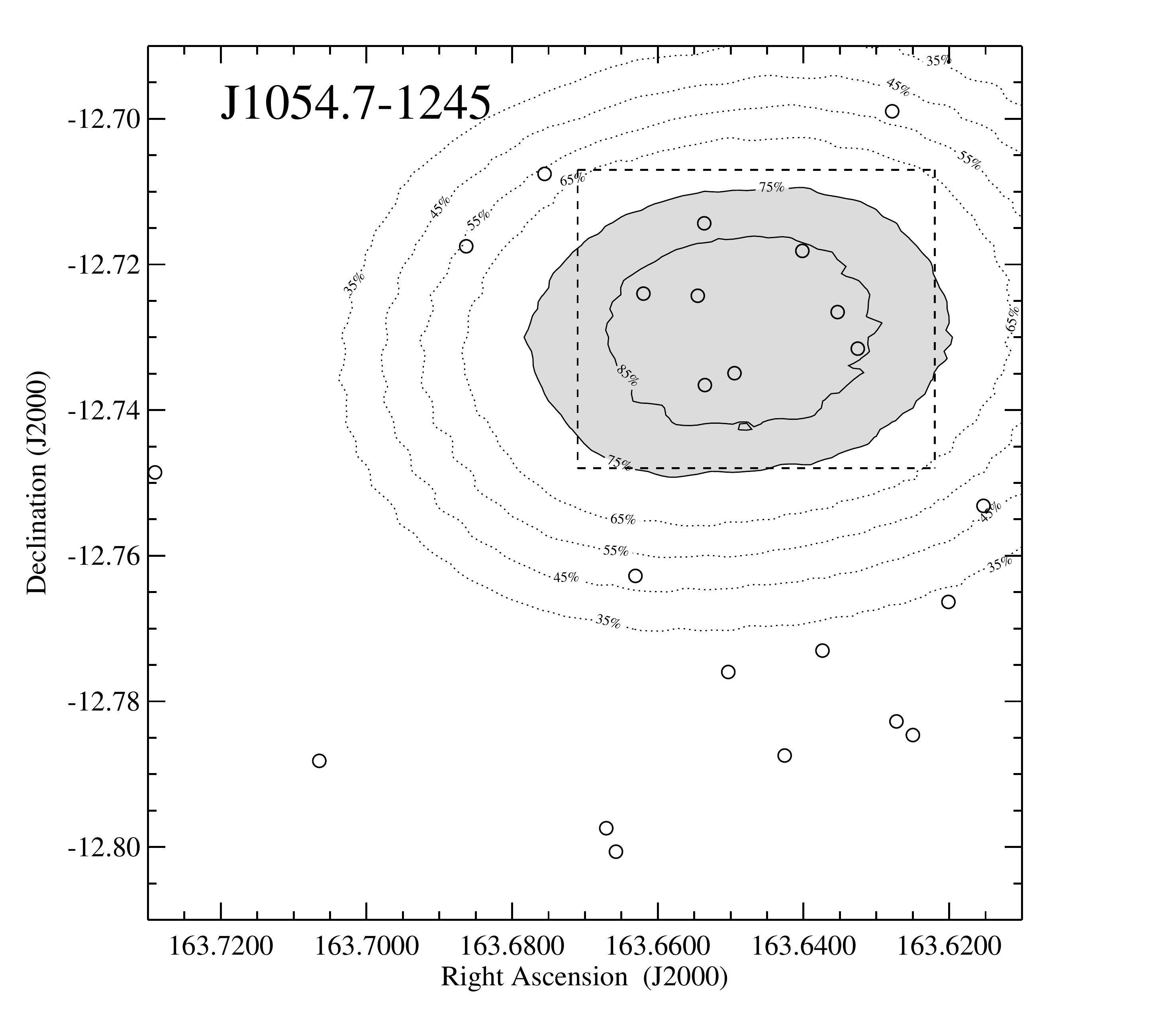}
\caption{Left: Over-density of LBGs in field J1054.7-1254 from the ERGS survey \citep[See][]{Douglas09}. Right: Coverage of MAMBO-2 observations, contours display level of coverage in comparison to the best coverage in the field. Shaded area indicates region covered by MAMBO-2 observations to $\sim$75\,\% of the best MAMBO-2 coverage (for which 7\,mm and 12\,mm observations are in hand). LBG positions are marked as circles, with their redshifts,  and the dashed box indicates region displayed in Figure \ref{fig:mambo_map} }

\label{fig:LBG_pos}
\end{center} 
\end{figure*}

\section{Observations}
\label{sec:obs}

Observations were undertaken using the MAMBO-2 array \citep{Kreysa98}
on the IRAM 30m telescope in the Spanish Sierra Nevada. MAMBO-2 is 117
channel bolometer with individual channels arranged in a circular
pattern.  It has a bandwidth centre of $\sim$250\,GHz (1.2\,mm) and
half power spectral range of 210 to 290\,GHz. The effective beam size
is 10.7$^{\prime\prime}$ ($\sim70$\,kpc at $z\sim4.95$) and it has an
undersampled field of view of 4$^{\prime}$. Field J1054.7-1245 was
observed on two separate occasions using On-the-Fly mapping for a
total integration time of 2.2 hrs on 2009 March-15/16\,th and 5.5 hours
during 2011 January-March. Each map was comprised of 18 azimuthal
subscans of 66\,s each, scanning at 5$^{\prime\prime}$s$^{-1}$, with
8$^{\prime\prime}$ steps in elevation between subscans. Primary
pointing, focus and flux calibrations were carried out on Saturn and
secondary pointing calibrations on nearby source J1058+016. Sky dip
observations were taken to correct for atmospheric opacity and
Precipitable Water Vapour (PWV) at the site was typically 0.35-0.7mm
(Tau CSO at 225GHz $\sim$0.1-0.4). The map was centred on the
most densely clustered region of the field for which 7mm and 12mm
observations are already in hand \citep[see][]{Davies10}. A total area
of $\sim5$ arcmin$^{2}$ was surveyed to 75\% of the best total
coverage encompassing eight LBGs (Figure \ref{fig:LBG_pos}). Data were
reduced using the dedicated MAMBO-2 reduction package MOPSIC (Version
19.01.2010, written by R. Zylka at IRAM) utilising an algorithm
designed to identify weak sources. Individual channels are corrected
for opacity and instrumental noise, and flux calibrated using data
obtained during the observations. A Correlated Signal Filter (CSF) was
used to reduce sky noise, a process designed specifically for the
detection of compact ($<0.5\, \times$ the beam size), weak
($\lesssim5\,\times$ rms) sources and which removes diffuse background
sources of flux. In the resultant map, a root mean square (rms) noise
of 0.82\,mJy/beam is obtained over the central $\sim5$ arcmin$^{2}$
region.

\section{Results}
\label{sec:results}

 \begin{figure}
\begin{center}
\includegraphics[scale=0.28]{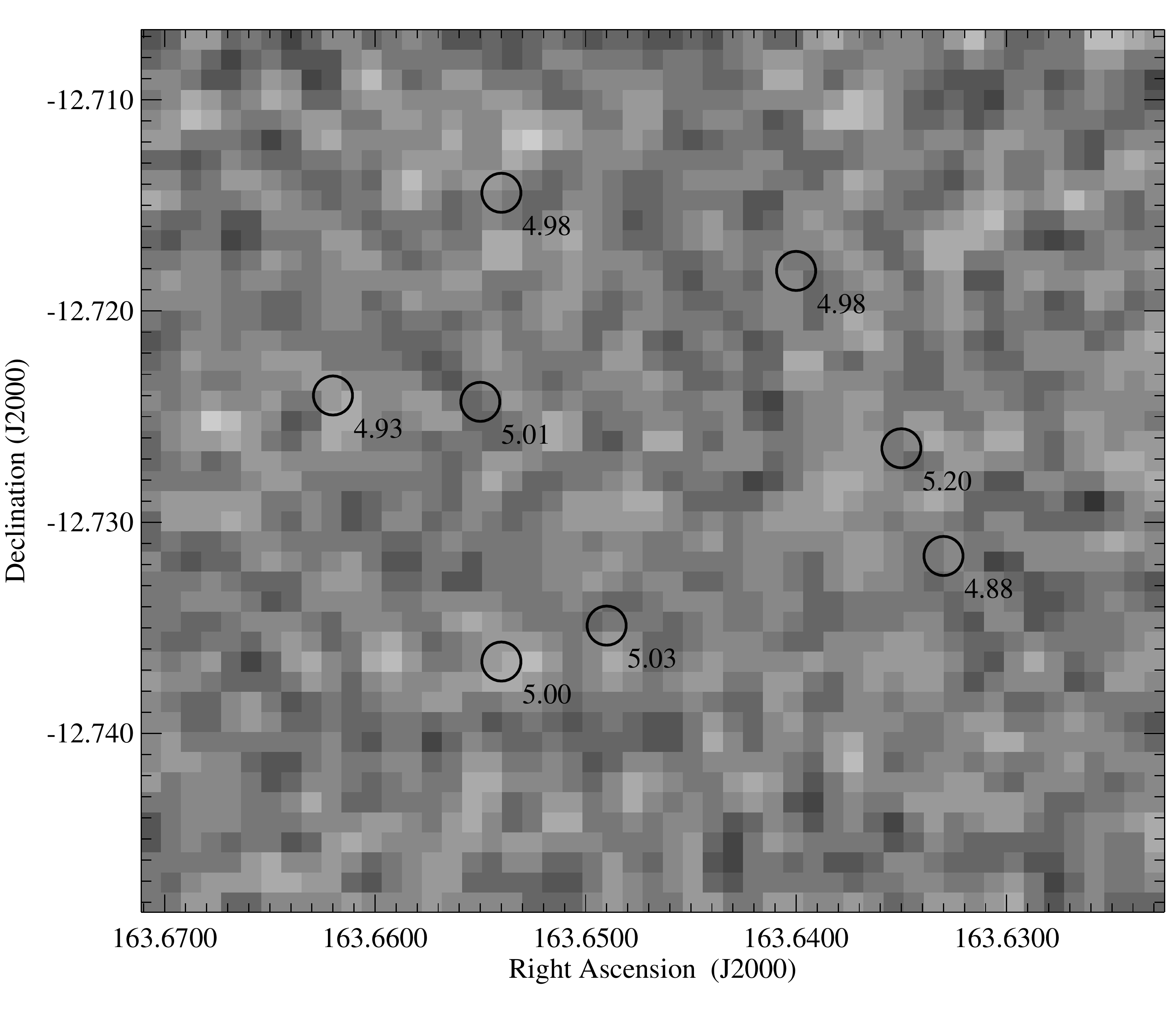}

\begin{center}
\hspace{17mm}
\includegraphics[scale=0.45]{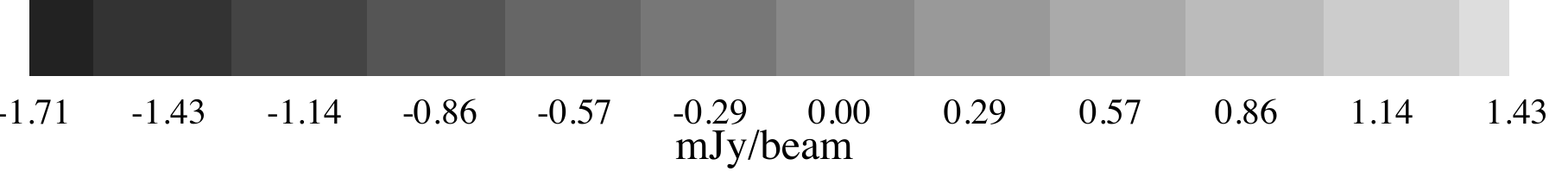}
\end{center}

\caption{MAMBO-2 map of the field (boxed region figure \ref{fig:LBG_pos}) encompassing eight LBGs at $z\sim5$. Positive flux is indicated as light regions. No individual source is detected at a 1.6\,mJy/beam (2 $\times$ rms) limit.}
\label{fig:mambo_map}
\end{center} 
\end{figure}

Eight spectroscopically-confirmed $z\sim5$ LBGs fall within the
central $\sim5$ arcmin$^{2}$ of the MAMBO-2 observations (Figure
\ref{fig:mambo_map}); seven lie within a narrow $z=4.95\pm0.08$ range
while the eighth is at $z=5.2$. Other spectroscopically-confirmed LBGs
that fall outside of this central region were excluded due to the
increased noise in the MAMBO-2 image at larger radii (more than twice
the rms achieved at the centre of the field), Figure
\ref{fig:LBG_pos}. No individual LBG is detected at $>\,2\,\times$ rms
(see figure \ref{fig:mambo_map}). There are only 4 potential
detections at twice the rms in this central region, and none at $3\times$ rms, broadly consistent with expected statistical
fluctuations. The immediate fields of each of the four potential
detections and the eight LBGs were examined using our optical, near-IR
and SPITZER/IRAC imaging of the larger fields
\citep{Douglas09,Douglas10} in order to search for optical
counterparts and assess the potential for source confusion. Only one
of the four potential detections showed a possible counterpart that
could account for the signal in the MAMBO-2 data.  One of the LBGs had
a source within 5$^{\prime\prime}$ of its position detected in all
bands from $V$ through to 8\,$\mu$m which could have confused the
interpretation had that LBG been detected in the MAMBO-2 data.

The typical half-light radii of a UV-luminous stellar component of an
LBGs in our sample is $\sim$ 0.2$^{\prime\prime}$ \citep{Douglas09},
while the MAMBO-2 beam has a FWHM of 10.7$^{\prime\prime}$ and so covers not
just the UV-luminous regions of any LBG, but the immediate projected $\sim
30$\,kpc around it. It is therefore expected that any emission from
sources will be unresolved in the MAMBO-2 maps even if it is extended
beyond the UV-emission region, and a 2\,$\times$ rms upper limit on
the flux from an individual source can be constrained at
$S_{1.2\mathrm{mm}} < 1.6\,$mJy.

\begin{figure}
\begin{center}

\includegraphics[scale=0.28]{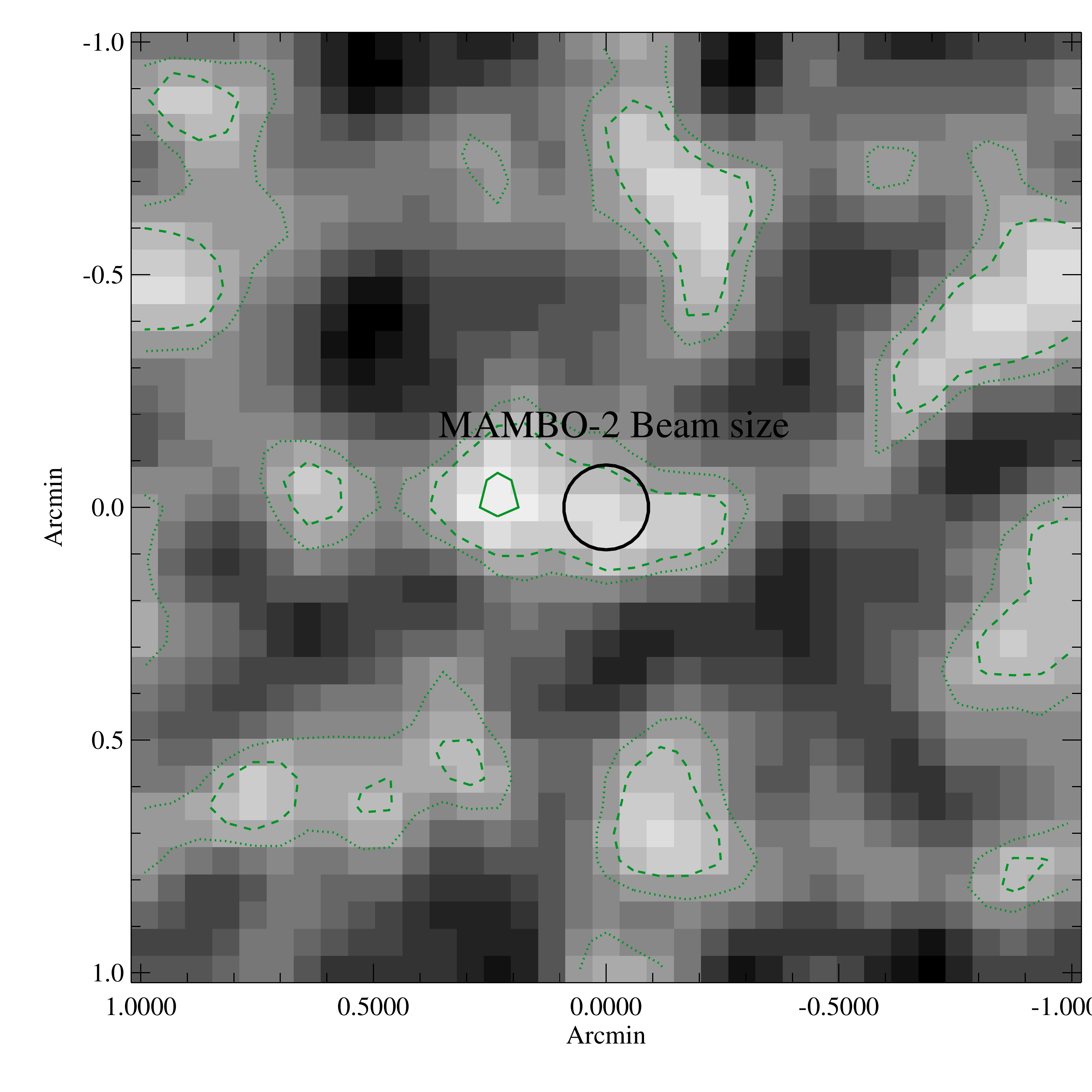}

\begin{center}
\hspace{12mm}
\includegraphics[scale=0.35]{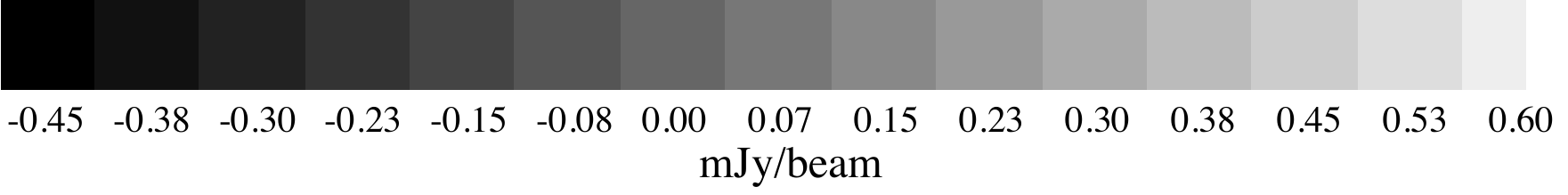}
\end{center}

\caption{Mean composite image of eight LBGs in the field with rms$\sim$0.29\,mJy/beam. Galaxies are stacked on the optically derived positions and image is convolved with the MAMBO-2 beam size (circle). Contours or 2, 1 and 0.5 sigma are displayed as solid, dashed and dotted lines respectively.}    

\label{fig:mean_stack}
\end{center} 
\end{figure}

In order to better constrain the flux from a typical LBG and its
immediate surroundings (within the beam FWHM), data were combined from
the positions of the eight LBGs. A $35\times35$ pixel
($\sim12\times12$ beam size) region centred on each LBG is extracted
from the map. These $35\times35$ pixel images are then combined into
an average image taking the mean of the eight images at every pixel
position. The resultant image was then convolved with a Gaussian
matching the MAMBO-2 beam size. The resultant composite image is shown
in Figure \ref{fig:mean_stack}.  Although there is a region of
positive flux at the centre of this composite image, it is extended by
at least two beam widths in one direction, with the peak not centred
on the position of the objects (beyond the expected pointing accuracy
of the telescope). Given the limited number of independent points in
this image (of order 100), the significance of this region is not
high.  We confirm this by repeating the stacking analysis using 8
random positions on the full MAMBO-2 image. The central $\sim5$
arcmin$^{2}$ region of this map contains only $\sim$\,255 independent
MAMBO-2 beam size regions. If 1000 realisations of a similar combined
image made from random combinations of the 248 positions that do not
contain a LBG are produced, it is found that $\sim18\,\%$ of the
realisations contain a $\gtrsim$ 0.6\,mJy/beam peak in the central
position (the flux in the central pixel of Figure \ref{fig:mean_stack}
is 0.6\,mJy/beam).  Consequently, we treat this flux as an upper limit
to the mean emission of these LBGS, so $S_{1.2\mathrm{mm_{MEAN}}} <
0.6\,$mJy, $\sim 2.5$ times lower than the $2\,\times$ RMS limit on
individual sources. 

The MAMBO-2 beam size
(10.7$^{\prime\prime}$) means that this limit not only
applies to the FIR flux arising from the rest-frame
UV-bright regions, but also includes any other
emission within $\sim 5^{\prime\prime}$ or $\sim 30$\,kpc (and certainly within $1^{\prime\prime}$)
of the sources. This flux may arise from either neighbours at the same redshift without
significant UV emission, or any underlying system within which
individual LBGs are embedded as UV-bright star forming
regions (see discussion). 

Although the MAMBO-2 field also encompasses the positions of several
spectroscopically-unconfirmed $z\sim 5$ LBGs candidates, we do not
include these in our composite analysis as if only a small number of
these sources are actually at $z\sim1-2$ \citep[which can show similar
photometric colours to $z\sim5$ LBGs, see][]{Stanway08a}, they could
bias an estimate for any FIR flux from true $z\sim5$ sources. For
completeness we also produced a composite image including the LBGs
which were excluded as having more than twice the rms achieved at the
centre of the field. This composite image did not improve upon the
depth obtained using the eight central LBGs - hence we do not discuss
it in any further analysis.

 \subsection{Joint constraints on L$_{\rm FIR}$, T$_{\rm dust}$ and $M_{\rm dust}$}             
 
\begin{figure}
\begin{center}

\includegraphics[scale=0.44]{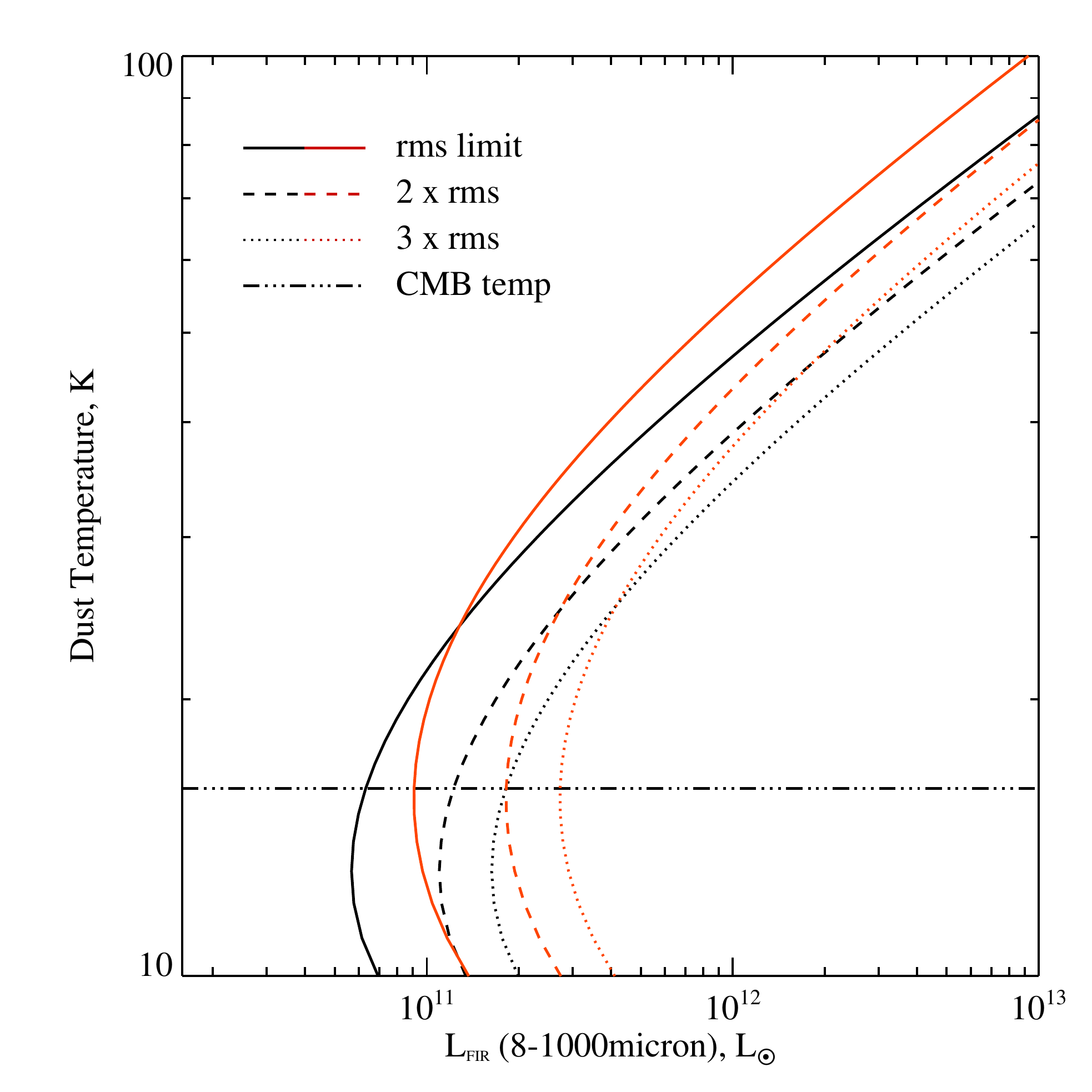}

\caption[Constants on the FIR Luminosity of a typical LBG at $z\sim5$.]{Constants on the L$_{\mathrm{FIR}}$ of a typical LBG at $z\sim5$ from our stacking analysis of the 1.2mm flux from eight galaxies (black); compared to limits derived from 870\,$\mu$m observations of similar sources \citep[red, ][]{Stanway10}. Solid lines display the L$_{\mathrm{FIR}}$ as a function of temperature for the rms limits. Dashed and dotted lines show the 2\,$\times$\,rms and 3\,$\times$\,rms limits. The temperature below which dust heating by the CMB occurs is plotted as a dot-dashed line.}    

\label{fig:L_FIR}
\end{center} 
\end{figure}

 In the absence of a detection, we can only jointly constrain the FIR
 luminosity and dust temperature of a typical $z\sim 5$ LBG, and
 therefore can only explore a limit on one if we make a reasonable
 assumption about the other. In order to relate the limiting flux to a
 joint constraint, we assume that any FIR emission has a grey-body
 spectrum arising from emission by dust heated by
 starlight. Accordingly we follow the procedure used in \cite{Aravena08} and 
 \cite{Stanway10}. We calculate the integrated FIR luminosity
 (L$_{\mathrm{FIR}}$) between 8-1000$\mu$m by integrating the grey
 body spectrum assuming an absorption coefficient of $\kappa_{\nu}
 \propto \nu^2$, as this provides the best fit to dust absorption in
 the spectra of high redshift quasars \citep[][]{Priddey01}. We
 normalise this assuming $\kappa=0.4$ cm$^2$ g$^{-1}$ at 250\,GHz
 \citep[See][]{Kruegel94}. In Figure \ref{fig:L_FIR}, the integrated
 FIR luminosity as a function of dust temperature is displayed for the
 flux limit obtained in the composite image.

Given the lack of previous study and detections of high redshift LBGs
in the mm/sub-mm, little is known about the appropriate dust
temperatures for such sources. \cite{Baker01} use SED fitting of a
single, highly lensed LBG at $z=2.7$ to obtain a dust temperature of
T\,=\,33\,K, and the mean dust temperature of submillimeter selected
galaxies at $z\sim2$ is $\sim35$\,K \citep{Chapman05}. We note
  that lower redshift ($z<4$) ULIRGS are found to have higher
  temperatures \citep[$\sim45\,$K, ][]{Rieke09}. Although such sources
  have very different redshifts and luminosities to the LBGs
  considered here, they demonstrate that higher temperatures are
  possible.

  In the absence of an exact dust temperature for these sources, we assume 30\,K as this is close to the temperature
  derived for the $z=2.7$ LBG - likely to be the most appropriate of
  the temperatures noted above. Using this, the FIR flux limit derived
  from the composite analysis of a typical z\,=\,5 LBG in this work
  gives a formal 2 $\times$ rms limit to the FIR luminosity of
  L$_{\mathrm{FIR}} \lesssim 3 \times 10^{11}$\,L$_{\odot}$. This
  limit is several times fainter than the FIR luminosities of $z\sim
  3$ galaxies detected in the sub-mm \citep[][]{Chapman05,Chapman09,
    Negrello10, Conley11}, which generally have L$_{\mathrm{FIR}}
  >10^{12}$\,L$_{\odot}$. With the caveat that the temperature is
  assumed rather than determined, the flux limit implies that typical
  $z\sim 5$ LBGs have FIR luminosities below those of $z<4$ ULIRGS and
  intermediate redshift BzKs \citep[$e.g.$ see][]{Daddi10}.

  While assuming a higher temperatures will lead to a higher FIR
  luminosity, it better constrains the associated dust mass, which
  varies inversely with temperature for a given flux limit. In order
  to place a temperature-dependent limit on the dust mass, we use the
  relationship between redshift-dependent flux density and dust mass
  outlined in equations 1 and 2 of \cite{Weib07}. This includes the
  effect of the increased Cosmic Microwave Background (CMB)
  temperature at high redshift, whereas the temperature of a source
  approaches that of the CMB, the mass required to distinguish it from
  CMB fluctuation increases dramatically. In this we assume a source
  size of 5 kpc, though the result does not change significantly for
  sizes between 2-5kpc \citep[typical sizes of $z\sim5$ LBGs, $e.g.$][]{Douglas09}.

\begin{figure}
\begin{center}

\includegraphics[scale=0.4]{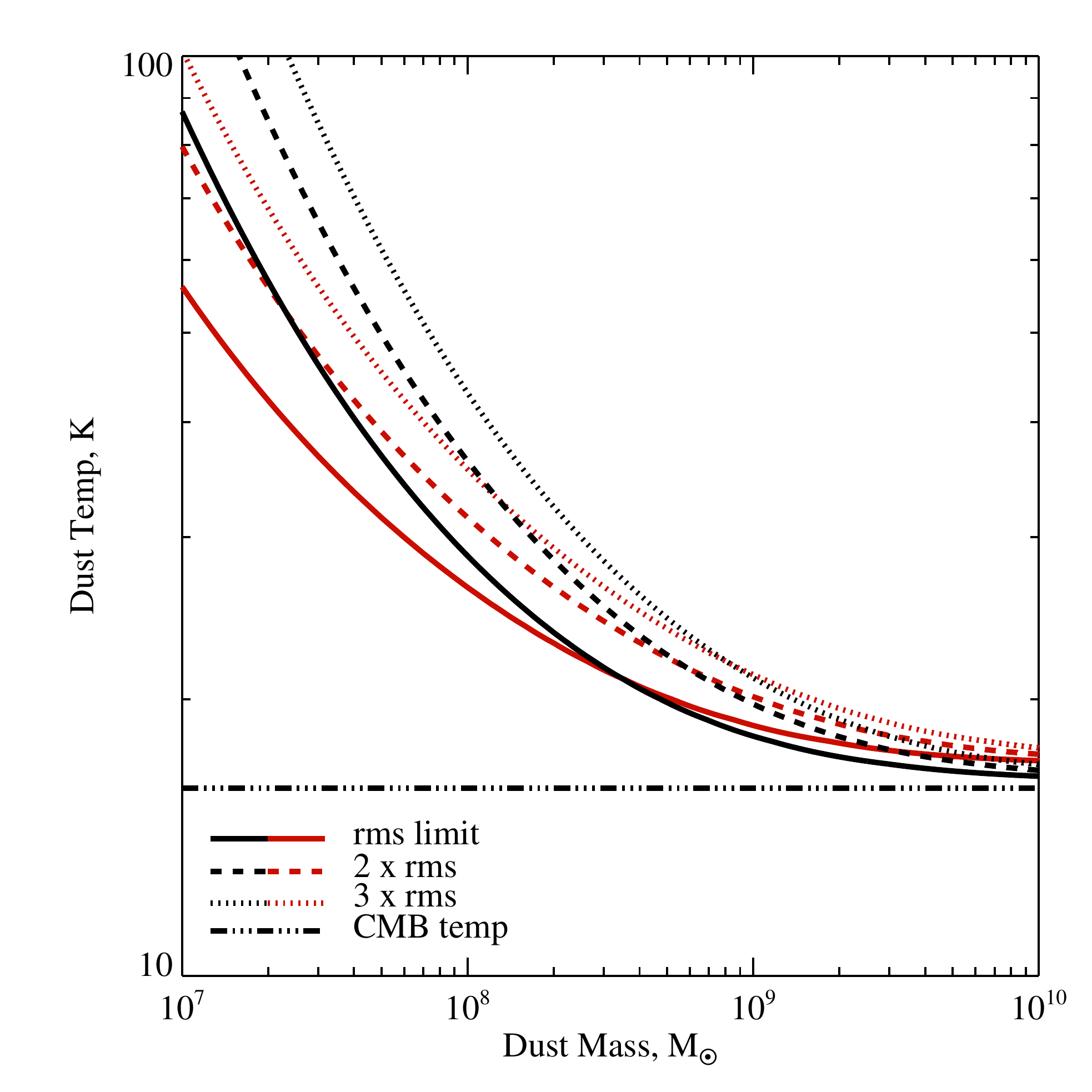}

\caption[Constants on the dust mass of a typical LBG at $z\sim5$.]{Constants on the dust mass of a typical LBG at $z\sim5$ from our stacking analysis of the 1.2mm flux from eight galaxies (black); compared to limits derived from 870\,$\mu$m observations of similar sources \citep[red, ][]{Stanway10}. Solid lines display the dust mass as a function of temperature for the rms limits. Dashed and dotted lines show the 2\,$\times$\,rms and 3\,$\times$\,rms limits. The temperature below which dust heating by the CMB occurs is plotted as a dot-dashed line.}    

\label{fig:dust_mass}
\end{center} 
\end{figure}

In Figure \ref{fig:dust_mass}, the dust mass as a function of dust
temperature is displayed for the flux limit obtained in the composite
image. To constrain the dust mass of a typical source in our field, an
assumption about the dust temperature must once again be made. A dust
temperature of 30\,K gives a formal 2 $\times$ rms limit to the dust
mass of M$_{\mathrm{dust}} \lesssim 10^{8}$\,M$_{\odot}$. At a higher
temperatures of 40\,K, approaching that found for lower redshift ULIRGS,
the mass limit is M$_{\mathrm{dust}} \lesssim 4 \times 10^{7}$\,M$_{\odot}$. Dropping the temperature to 20\,K leads to a far weaker
mass limit, M$_{\mathrm{dust}} \lesssim 4 \times 10^{8}$\,M$_{\odot}$. However this temperature is approaching that of the CMB
at $z\sim 5$. Given that there is clearly vigorous star-formation in
the LBGs, it seems unlikely the dust would remain unheated by it or be close to equilibrium with the CMB. Hence, the naive expectation
is that the dust temperature will be higher than this.

\subsection{Obscured star-formation rates}

Star-formation rates (SFRs) in typical $z\sim5$ LBGs have routinely
been determined through SED fitting to their rest-frame UV-optical
emission (sampling star-formation which is subject to little dust
obscuration), giving typical values of ${\rm few}\times
10\,$M$_\odot$\,yr$^{-1}$ \cite{Verma07}. While our non-detection in
the composite image is consistent with a significant proportion of
star-formation within $z\sim5$ LBGs being relatively free of dust extinction, it
is nonetheless interesting to consider the limit to the dust obscured
SFR imposed by our non-detection. Using the Kennicutt relation for
local starburst galaxies \citep{Kennicutt98} and our
L$_{\mathrm{FIR}}$ limit we obtain an obscured SFR limit of
$\lesssim52 $\,M$_\odot$\,yr$^{-1}$. This suggests that at least 40$\%$ of the total star-formation in $z\sim5$ LBGs is unobscured.

\subsection{Combining with LABOCA data}

In addition to investigating the limits derived from this study, we
can also combine our 1.2\,mm observations with LABOCA(870\,$\mu$m) imaging
of a similar, $z\sim5.16$ LBG over-density from \cite{Stanway10} in an
attempt to obtain a limit to or detect the combined emission from the
17 spectroscopically-confirmed $z\sim 5$ LBGs across the two data
sets. We do this by combining the average stacked 1.2mm image with a
similar average stacked 870\,$\mu$m image obtained by
\cite{Stanway10}. We take into account the different beam sizes and
pixel scales, scaling the 870\,$\mu$m limit to that at 1.2mm by assuming
a temperature and spectrum of 30K grey body for the average source and
accounting for different mean redshifts of the two LBG samples.  No
source is identified at the central position in this combined image at
a 2 $\times$ rms limit of $\sim$0.44\,mJy/beam at 1.2mm, representing
a limit on the average flux of the 17 LBGs.

\section{Discussion}
\label{sec:discuss}

\subsection{Non-detection in the composite image and implications for $z\sim5$ sources}

\subsubsection{Model predictions for FIR emission $z\sim5$ LBGs}

Until recently, there have been no self-consistent model-based
predictions for the expected fluxes of sources such as these. After
these observations were carried out, \cite{Gonzalez11} made
predictions for the 850$\mu$m flux of distant LBGs as a function of
their UV luminosities, based on the GALFORM semi-analytic model of
galaxy evolution. Although they concentrate on redshifts $z=3$ and
$z=6$, the small variation in prediction at around at rest-frame UV absolute magnitude, M$_{1500(AB)}-$5log(h) $\sim -19.8$ (the typical value for the
$\sim 1500$\AA\  absolute magnitude of the ERGS sources, where h is the dimensionless Hubble parameter) indicates we can use these predictions
at $z\sim 5$. Looking at their figure 11, the median flux predicted is
around 0.4\,mJy at 850\,$\mu$m, which we calculate to be equivalent to
0.2\,-\,0.25\,mJy at 1.2\,mm (for grey body temperatures in the range
30\,-\,50\,K). 

We also infer from their figure 11 that about ten per cent of
their sources should be brighter than $\sim 1\,$mJy at 850\,$\mu$m. The
limits determined in both this and the earlier work of
\cite{Stanway10} for the typical source, are roughly double the median
value predicted by \cite{Gonzalez11}. While our non-detections are
compatible with the model, it is clear from our attempt at combining
both the current and \cite{Stanway10} data sets that a sample size
double that of the current one (observed to similar depths or deeper)
will start to challenge or validate these model predictions. Of course, this is just one model with one set of assumptions
and prescriptions; having the ability to test more models which use
different methodologies would increase the utility of such
observations. Interestingly an alternate model by \cite{Finkelstein09},
based on the assumption that sub-mm emission in these sources is
powered by the UV-luminous component, gives predictions that span a
range covering both the predictions of \cite{Gonzalez11} and our limits on
the typical LBG.

\subsubsection{The nature and baryonic content of $z\sim5$ LBGs}

As noted earlier, we obviously cannot know the dust temperature for
any material within these galaxies in the absence of detections at
multiple frequencies. However, by assuming a temperature we are able
to explore how the limit to the dust mass relates to the mass in stars
for these sources. For temperatures of 30\,K or above, a dust mass
limit for LBGs at $z\sim5$ is found to be roughly $1\%$ to $10\%$ of
the typical mass of their stellar component \citep[depending on study,
for a typical $z\sim5$ LBG similar to those observed here, M$_{\star}
\sim$ few $\times10^{9-10}$\,M$_{\odot}$ at least,][]{Verma07,McLure10}. In
addition, the molecular gas content of a typical LBGs in this field has been
found to be comparable to this typical stellar mass, at most
\citep[M$_{\mathrm{H2}}$\,$<$ a few $\times10^{9}$ M$_{\odot}$,
assuming a CO flux to H$_{2}$ mass conversion `X-factor'\,=\,0.8\,M$_{\odot}$(K\,km\,s$^{-1}$\,pc$^2$)$^{-1}$ for
consistency with earlier work, see][ and references therein]{Solomon05,Davies10}. Hence, assuming a dust temperature of $>30$K, no more than $\sim 50\%$ of the total
baryonic mass of LBGs at $z\sim5$ is in their dust and molecular gas
component - to the current limits and given the assumptions above. Clearly, using a larger CO flux to H$_2$ mass conversion factor will
increase the upper limit to the molecular gas content. Using an
X-factor more typical of low redshift spirals, 4.5\,M$_{\odot}$\,(K\,km\,s$^{-1}$\,pc$^2$)$^{-1}$, with these limits, allows up to $\sim80\%$ of the baryonic mass of $z\sim5$ LBGs to be in the
form of molecular gas and dust if a typical source has a stellar mass
of $\sim 10^9$M$_{\odot}$. For sources with stellar masses closer to
$10^{10}$M$_{\odot}$, even with this X-factor, the gas would account for less
than $50\%$ of the baryonic mass. As we know little of the correct CO
flux to H$_2$ mass conversion factor for these sources we can not constrain this fraction further and must consider this limitation in any
conclusions drawn from such an analysis.

By determining limits to the total baryonic content of $z\sim5$ LBGs
we can explore the nature of the LBG phenomenon at these redshifts.  A
substantial fraction of these objects show distorted or multiple UV
components on sub-arcsecond scales, corresponding to a few kpc
\citep[e.g.][]{Concelise09, Douglas10}. There has been detailed
discussion in the literature as to what these morphologies tell us
about the nature of LBGs. Through a comparison to low redshift LBG
analogues, \cite{Overzier08} show that such morphologies can be
explained as the result of gas rich mergers of low mass systems
triggering starbursts, while \cite{Law07} explore the possibility that
the UV components are relatively unobscured regions of star-formation
in a larger underlying galaxy which is either intrinsically less
UV-luminous or is extincted.

Given the separation between components (typically
$\sim0.5^{\prime\prime}$, a few kpc), any encompassing or underlying
galaxies clearly have to be larger and more massive than the sizes and
masses attributed to the UV-luminous regions (so with a total stellar
mass $>10^{10} $M$_{\odot}$ and possibly considerably larger). Both a
gas rich merger or a larger underlying system could easily lead to
appreciable far infra-red (FIR) luminosities, through significant
rapid dust production in the case of the merger or simply by the
amount of obscuring material/total mass in the case of the underlying
system. However, it is perhaps easier to explain a lower FIR
luminosity in the case of a merger, as it could be argued that in
these LBGs the UV emission traces the majority of the material in the
system.

The apparent lack of large quantities of material that could be
detected in the millimetre and sub-millimetre (dust and molecular gas)
in these systems weakens the case for LBGs being unobscured super
starburst regions embedded in much larger systems, \citep[see
discussion in][]{Davies10}. While this also argues against very
gas-rich mergers that promptly produce large amounts of dust, it poses
no constraint on mergers where most of the baryonic material is traced
by the UV emission and is in the form of stars.

\subsection{Comparisons to lower redshift populations and implications for future observations}

Given the lack of detections in the current data it is worthwhile
exploring how much deeper further observations may need to probe
before the population is detected in emission from dust. We can do
this by investigating the properties of lower redshift ($z<4$) sources
which, while not ideal, may have use as  analogues for $z\sim5$
LBGs.

The limit on the stellar mass to cold dust ratio obtained here
($>$\,10) is similar those observed in some local starbursts. In a study of
metal poor blue compact dwarf galaxies, \cite{Hunt05} derive stellar
to dust mass ratios of 7.5 and 13 for two sources. Both stellar and
dust masses in \cite{Hunt05} are derived from SED fitting and no
errors are given, hence these are only best-fit values. Although these
sources are less massive than LBGs at $z\sim5$, they have comparable
specific star-formation rates and metallicity (0.14\,Z$_{\odot}$ and
0.2\,Z$_{\odot}$ compared to 0.1\,-\,0.2\,Z$_{\odot}$ obtained for
$z\sim5$ LBGs from SED fitting, $e.g$ Verma et al. 2007, and
rest-frame UV spectral slope analysis, Douglas et al. 2010). In
addition, these local starbursts have emission dominated by
small-grain, Type-II SNe formed dust as is present in high-$z$ sources
\citep{Maiolino04} and therefore, they should display similar dust
characteristics.

If these systems are good models for
$z\sim 5$ LBGs, observations not much deeper than those presented here
should start to detect the typical LBG at FIR
wavelengths. For example, using a stellar mass of M $=10^{9}$\,
  M$_{\odot}$ and dust temperature of $>30$\,K, a stellar to dust mass
  ratio of $<13$ predicts a dust mass of M$_{\mathrm{dust}} \lesssim 8
  \times 10^7$\,M$_{\odot}$. Assuming the same grey body model outlined
  above, an rms of $\sim0.14$\,mJy is required for a 2\,$\sigma$ detection
  of such a source. For a similar MAMBO-2 observation and stacking the
  8 LBG in the field a $\sim 30$h of integration time are required to
  reach this limit. In comparison, for a similar integration time to that in this study, a
  composite image of $\gtrsim30$ LBGs would be required to obtain a
  detection. 
 
  A more direct comparison can be made with UV-selected galaxies at
  lower redshift. LBGs at $z\sim3$ are identified via bright
  rest-frame UV continuum emission using a similar method to those at
  $z\sim5$. Their UV/optical properties indicate they have comparable
  star-formation rates, but with higher metallicities, slightly larger
  dust extinctions in the rest-frame UV and have typically older and
  more massive detectable stellar population (by close to an order of
  magnitude) than $z\sim5$ LBGs \citep[see][]{Verma07}, so their use
  as analogues is limited.
  
\cite{Rigopoulou10} detect the average emission from IR-luminous
  $z\sim 3$ LBGs at 250$\mu$m with {\it Herschel-}SPIRE and calculate
  L$_{\rm FIR}=2.8\times10^{12}\,$L$_\odot$ for T$_{\rm
    dust}=45$K. Looking at figure 4, we note that if we assumed this
  temperature, our flux limit already probes this luminosity at $z\sim
  5$. However, for sources not selected to be IR-luminous, the picture
  is different. Thermal dust emission has been detected in several
  strongly lensed LBGs at $z\sim3$ giving typical dust masses of
  $\sim$ few $\times \, 10^{7}$M$_{\odot}$
  \citep[e.g.][]{Baker01,Coppin07}.  \cite{Stanway10} calculate a
  similar limit to the dust mass of typical $z\sim 3$ LBGs from the
  flux limit obtained by \cite{Webb03} from observations of such
  sources in the Canada-UK Deep Submillimeter Survey (again assuming
  $T_{\rm dust}=30$K).  If these values are also appropriate at higher
  redshift, then significantly deeper observations will be required to
  detect the dust emission from LBGs at $z\sim5$, observations which
  may run into the confusion limit of single-dish observations.
  
\begin{figure}
\begin{center}

\includegraphics[scale=0.4]{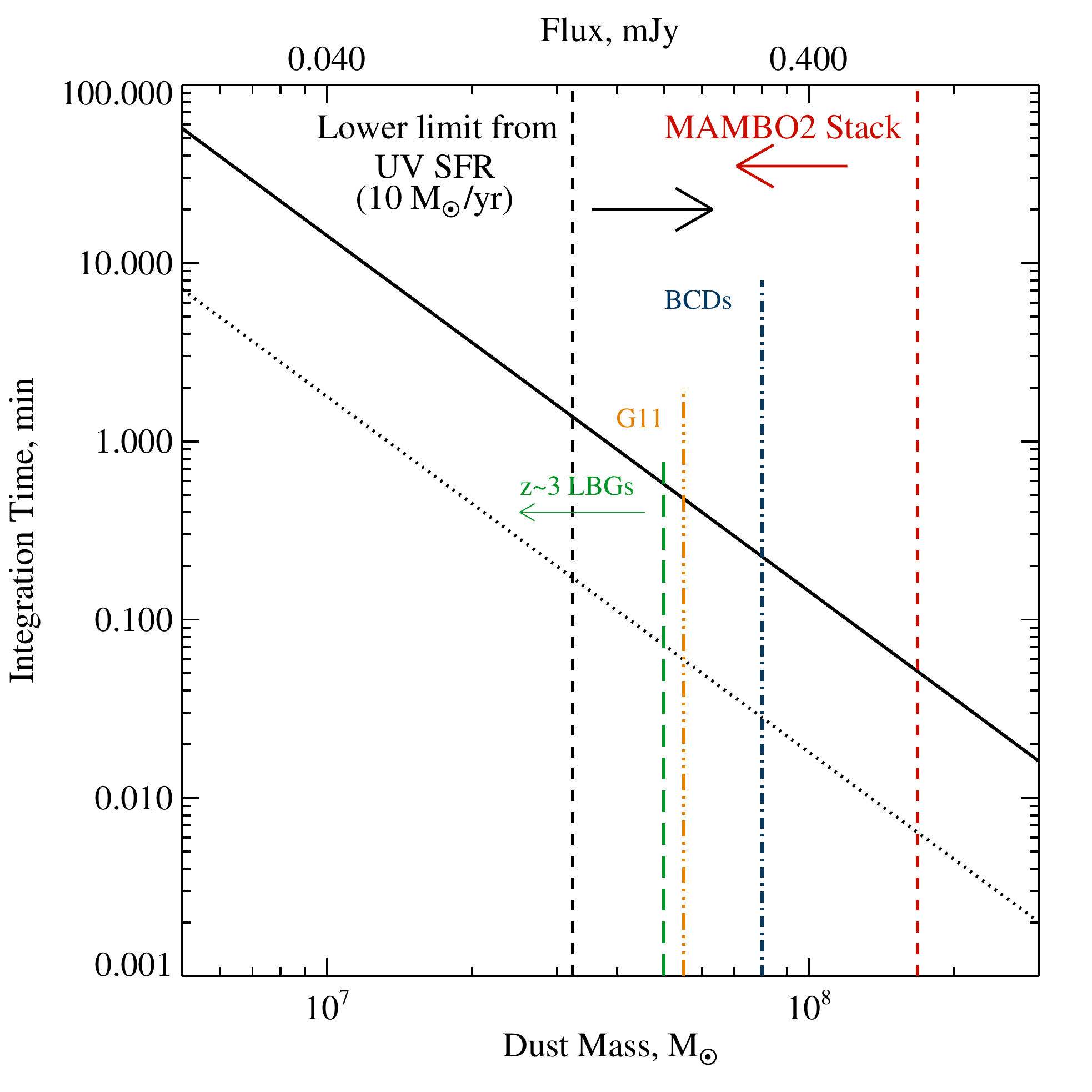}

\caption{The detection limits obtainable as a function
  of integration time for the full sensitivity (50 antennae) ALMA
  array. The ALMA exposure time calculator was used to determine the
  $2\times$rms for a range of exposure times. This was converted to a
  dust mass limit assuming a 30\,K grey body as in the rest of this
  work. The solid diagonal line displays limits achievable for a
  single source and dotted line the limits achievable when stacking 8
  sources, as carried out here. The red vertical dashed line indicates
  the mass limit derived from these MAMBO-2 observations. The blue
  dot-dashed line displays the predicted dust mass if $z\sim5$ LBGs
  have the same M$_{\star}$/M$_{dust}$ as blue compact dwarf galaxies
  \citep{Hunt05}.  The orange dashed-dot-dot line displays the
   \citet{Gonzalez11} predictions for $z\sim5$ LBGs. The green large
   dashed line displays the dust mass limit calculated from the
    \citet{Webb03} sample of $z\sim3$ LBGs. The black vertical dashed
   line is a predicted $lower$ limit determined by applying the Kennicutt
  relation to a star-formation rate of $\sim10\,$M$_\odot$/yr, typical
   for the UV-detected star-formation rates in $z\sim5$ LBGs, and
  converting to a dust mass assuming a T$_{dust}$=30\,K. }

\label{fig:alma_times}
\end{center} 
\end{figure}

  However, the required depth will be straightforwardly
  achievable with ALMA when it reaches full science mode. Figure
  \ref{fig:alma_times} displays the dust mass limit as a function of
  integration time for ALMA in full science mode assuming a grey body
  at 30\,K. These values are calculated using the ALMA exposure time
  calculator observing a field at a declination of -12 deg at
  250\,GHz, with 7.5\,GHz bandwidth, T$_{sky}$=44.3\,K and
  T$_{sys}$=132.5\,K. Clearly ALMA will make the detection of FIR
  emission from such sources routine, obtaining the same limit as our
  MAMBO-2 observations in $<10$ seconds and detecting
  M$_{\mathrm{dust}} = 10^7$\,M$_{\odot}$ sources at z$\sim5$ in just
  tens of minutes. While this represents a relatively short exposure
  for ALMA, it does mean that these sources will have to be targeted
  individually rather than as part of anything other than a deep blind
  survey. The source density of spectroscopically-confirmed ERGS LBGs
  is approximately one per 6 arcmin$^2$, or one source per $\sim 50$
  ALMA primary beams at 250\,GHz.  While fainter LBGs are more
  numerous, they are also likely to be fainter in the mm/sub-mm
  \citep[$e.g.$ ][]{Gonzalez11}.
        
\subsection{Overlap between LBG and SMG populations at $z\sim 5$}

Several spectroscopically-confirmed $z>4$ SMGs have been identified in
the past three years \citep[{\it e.g.} see the list in
][]{Kohno11}. While the optical photometry of these sources is
incomplete in the literature, of the seven sources in \cite{Kohno11},
five would have been selected in a survey using the Lyman break
technique \cite[see][]{Schinnerer08,Coppin09,Daddi09a, Daddi09b, Riechers10, Knudsen10}. Given that a large fraction of distant SMGs are potentially
detectable in LBG surveys \citep[even if for some of them the UV-light
is contaminated by AGN emission, unlike most LBGs, e.g.][]{Coppin09}, it is
informative to quantify the overlap between the population of typical
$z\sim 5$ LBGs and the more extreme (as far as star-formation rates
are concerned) SMGs at this redshift.

We have shown that the typical $z\sim 5$ LBG is not a
strong FIR source. By combining our results with those of
\citet[][]{Stanway10} we can place a limit on the fraction of LBGs
that could be mJy sources at $\sim 1$\,mm. None of the 17
spectroscopically-confirmed sources probed are detected individually
at 1-3\,mJy and together have an average flux of less than 1\,mJy
implying that less than 10\% of these confirmed LBGs have
submillimeter flux at the milli-Jansky level.  In addition to these
sources, both clustered fields contain multiple
photometrically-selected LBG candidates without spectroscopic
confirmation, either because they were not targeted by spectroscopy
due to observational constraints, or because spectroscopy was
inconclusive \citep[][]{Douglas09,Douglas10}. None of these were
detected at 870\,$\mu$m or 1.2\,mm.  Accounting for the fraction of
these candidates that will be at $z\sim 5$ \citep{Douglas09}, this almost
doubles the number of $z\sim 5$ LBGs covered in the deep regions of
the two fields, indicating that less than 5\,\% of $z\sim 5$ LBGs have
flux at 2\,-\,3\,mJy.

In order to place an upper limit on the fraction of FIR bright sources
in the $z\sim 5$ LBG population, we can extrapolate further. Given the
volume probed and the lack of detections in both this and the
\cite{Stanway10} work, if the typical source is detected at just below
our limits (this does not have to be true but will maximise the
fraction of bright sources) and if at this flux level we probe the
exponential (bright) end of the LBG FIR luminosity function, we expect
less than 1\,\% of the spectroscopically confirmed $z\sim 5$ LBG
population to give a $>5$\,mJy detection. If the typical source is
considerably fainter, then an even smaller fraction will be this
bright. 

Furthermore, it should be remembered that these sources are
selected from the two most over-dense fields of the ten similarly
sized ERGS pointings. If there is any bias towards more massive
starburst galaxies being preferentially found in  the regions with the most
evolved large scale structure at high redshift \citep[the most massive
systems at any epoch are likely to be found in the most over-dense
environments, $e.g.$ see][]{Mo02, Blain04}, the lack of a detection in
these fields may indicate an even lower limit on the fraction of
potential FIR bright LBGs in this and in any unbiased LBG survey as a
whole.

\subsection{The lack of other detections in the field}

Given the expected typical sub-mm flux of LBGs and known sub-mm number
counts, it is perhaps unsurprising that no sub-mm bright sources were detected
within this field to our limiting flux \citep[at $\sim700$
sources/deg$^{2}$, we would expect $\sim1\pm1$ source in a randomly
chosen field this size, based on the numbers in ][]{Hatsukade11}.  However, this field was
selected because of its unusual clustering of $z\sim 5$ LBGs, indicating that it contains a significantly over-dense, and
therefore evolved, structure at this redshift.  

It has been argued that luminous SMGs require significant early
hierarchical evolution of their environments to exist and hence should
be found in high density peaks of the matter distribution at high
redshift \citep[e.g.][]{Blain04}. Recently, a $z\sim 5.3$ SMG was
discovered within an over-density of photometrically-selected LBGs at
the same redshift \citep{Capak11}, supporting this picture. The LBG
over-densities in the currently studied field and in that studied by
\citet{Stanway10} are comparably dense. All three fields could
potentially be the progenitors of low redshift massive clusters, or at
least go on to form very massive galaxies by the
present-day. Therefore, they are equally plausible regions in which to
search for massive (UV-obscured) submillimeter galaxies.

Despite the LBG clustering indicating a high density environment, we
do not detect any sources within either field with fluxes above a few
mJy (with FIR luminosities of more than a few $\times
10^{11}$\,L$_{\odot}$ and therefore, dust masses of above a few
$\times 10^8$\,M$_\odot$ assuming $T=30$\,K, and inferred star
formation rates above a few$\,\times \,100$\,M$_\odot$\,yr$^{-1}$). A
possible explanation for this is that the FIR luminous stage giving
rise to SMGs is expected to be short ($\sim$100\,Myr). Therefore, it
is entirely possible that at least one such system, which is at an
evolutionary stage when it is not FIR luminous, will be in the
extended structure rendering it undetectable in our observations. The
typical timescale of the LBG phase at these redshifts is comparable or
less \citep[$\lesssim100$\,Myr,][]{Verma07,McLure10}. However, as
discussed in \cite{Stanway08} and \cite{Douglas10}, given the
relatively high volume density of such sources, as some LBGs fade in
the UV, others ``switch on''. This allows the large-scale structure to
be continually traced by LBGs, albeit different LBGs at different
times. Given the lower volume density of SMGs it is possible that we
have simply observed the structure at a time when none of the possible
SMG systems are FIR luminous.

These MAMBO-2 observations have not probed the entire
spatial extent of the over-density identified in the optical data (see
Figure 1) and it is possible that an SMG similar to that identified in
\citet{Capak11} could lie beyond the high sensitivity region in Figure
1. However, the observations are centred on the most over-dense region
in the field, perhaps the most likely position for such an
object. Obviously the sensitivity of these observations is such that
an object with a FIR luminosity (and star-formation rate, dust mass
{\it etc}) a few times lower than the known $z\sim 5$ SMGs would not
have been detected. Given that the known objects represent the extreme
of the population, we cannot rule out this possibility.

\section{Conclusions}

Constraining the thermal dust component of LBGs at $z\sim5$ is an
important step in developing a picture of star-formation and galaxy
evolution at high redshift. We have carried out 1.2mm MAMBO-2
observations of a field over-dense in LBGs at $z\sim5$, allowing the
simultaneous observation of eight LBGs. No individual source is
detected at a 1.6\,mJy/beam (2 $\times$ rms) limit. When stacking the
flux from the positions of all eight galaxies we obtain a limit of
0.6\,mJy/beam for the average emission from the objects. Assuming a
dust temperature of $T=30$\,K this corresponds to a FIR luminosity and
dust mass limit for a typical LBG at $z\sim5$ of L$_{\mathrm{FIR}}
\lesssim 3 \times 10^{11}$ L$_{\odot}$ and
M$_{\mathrm{dust}}\,\lesssim\,10^8$\,M$_{\odot}$, less than ten per
cent of the stellar mass ($\sim10^9$\,M$_{\odot}$). However, this dust
mass limit is  dependant on dust temperature, which is yet to
be constrained in high redshift galaxies. We estimate a limit to the total star-formation
rate in a typical $z\sim 5$ LBG of $\lesssim 52$M$_\odot$\,yr$^{-1}$
using the Kennicutt relation. This compares to minimum star-formation
rates of a few $\times 10$M$_\odot$\,yr$^{-1}$ determined from the
rest-frame UV emission of typical $z\sim 5$ LBGs. When combined with
our previous results on a similarly clustered field \citep{Stanway10},
we have observed 17 spectroscopically-confirmed $z\sim 5$ LBGs with
uniform results, ensuring that these studies robustly characterise
this population.

No other objects are
identified in the field, ruling out any source similar to the known
$z\sim 5$ SMGs lying in the same, highly over-dense structure as the
observed LBGs.

The  limits on the dust mass within the LBGs and their immediate
($\sim 30$kpc) environments weakens the possibility that these objects
are relatively unobscured super starburst regions embedded in
dust-obscured more massive, larger systems, especially when combined
with the limit on the molecular gas emission from our earlier work.

Between this and our previous work we have characterised the molecular
gas and dust properties of the typical population of $z\sim 5$ LBGs to
the limit of the available instrumentation. With the advent of ALMA we
should be able to straightforwardly detect both the cool gas and dust
phases of these LBGs and their environments (in targeted
observations), even if they prove to have a baryon content towards the
lower end of the predicted expectations.

\section*{Acknowledgements}

LJMD, EM and ERS acknowledge funding from the UK Science and Technology Facilities Council (STFC).


\begin{thebibliography}{}

\bibitem[Aravena et al.(2008)]{Aravena08} Aravena, M., Bertoldi, F., Schinnerer, E., et al.\ 2008, \aap, 491, 173

\bibitem[Baker et al.(2001)]{Baker01} Baker, A.~J., Lutz, D., Genzel, R., Tacconi, L.~J., \& Lehnert, M.~D.\ 2001, \aap, 372, L37 


\bibitem[\protect\citeauthoryear{Blain et al.}{2002}]{Blain02} 
Blain A.~W., Smail I., Ivison R.~J., Kneib J.-P., Frayer D.~T., 2002, PhR, 
369, 111 

\bibitem[\protect\citeauthoryear{Blain et al.}{2004}]{Blain04} 
Blain A.~W., Chapman S.~C., Smail I., Ivison R., 2004, ApJ, 611, 725 

\bibitem[Bouwens et al.(2007)]{Bouwens07} Bouwens, R.~J., 
Illingworth, G.~D., Franx, M., \& Ford, H.\ 2007, \apj, 670, 928

\bibitem[\protect\citeauthoryear{Capak et al.}{2011}]{Capak11} 
Capak P.~L., et al., 2011, Nature, 470, 233 



\bibitem[Carilli et al.(2010)]{Carilli10} Carilli, C.~L., et al.\ 
2010, \apj, 714, 1407 

\bibitem[Cazaux \& Spaans(2004)]{Cazaux04} Cazaux, S., \& Spaans, M.\ 2004, \apj, 611, 40


\bibitem[Chapman et al.(2005)]{Chapman05} Chapman, S.~C., Blain, 
A.~W., Smail, I., \& Ivison, R.~J.\ 2005, \apj, 622, 772 

\bibitem[Chapman \& Casey(2009)]{Chapman09} Chapman, S.~C., \& Casey, C.~M.\ 2009, \mnras, 398, 1615 

\bibitem[Cole et al.(2000)]{Cole2000} Cole, S., Lacey, C.~G., 
Baugh, C.~M., \& Frenk, C.~S.\ 2000, \mnras, 319, 168 


\bibitem[\protect\citeauthoryear{Conselice 
\& Arnold}{2009}]{Concelise09} Conselice C.~J., Arnold J., 2009, MNRAS, 397, 208 

\bibitem[Conley et al.(2011)]{Conley11} Conley, A., et al.\ 
2011, \apjl, 732, L35

\bibitem[Coppin et al.(2007)]{Coppin07} Coppin, K.~E.~K., et al.\ 2007, \apj, 665, 936 

\bibitem[\protect\citeauthoryear{Coppin et al.}{2009}]{Coppin09} 
Coppin K.~E.~K., et al., 2009, MNRAS, 395, 1905 

\bibitem[Coppin et al.(2010)]{Coppin10} Coppin, K.~E.~K., et 
al.\ 2010, \mnras, 407, L103 


\bibitem[Daddi et al.(2009a)]{Daddi09a} Daddi, E., Dannerbauer, 
H., Stern, D., et al.\ 2009a, \apj, 694, 1517 

\bibitem[Daddi et al.(2009b)]{Daddi09b} Daddi, E., Dannerbauer, 
H., Krips, M., et al.\ 2009b, \apjl, 695, L176 

\bibitem[Daddi et al.(2010)]{Daddi10} Daddi, E., Elbaz, D., 
Walter, F., et al.\ 2010, \apjl, 714, L118 

\bibitem[Davies et al.(2010)]{Davies10} Davies, L.~J.~M., 
Bremer, M.~N., Stanway, E.~R., Birkinshaw, M., 
\& Lehnert, M.~D.\ 2010, \mnras, 408, L31


\bibitem[Douglas et al.(2007)]{Douglas07} Douglas, L.~S.,  Bremer, M.~N.,  Stanway, E.~R.,   Lehnert, M.~D.\ 2007, \mnras, 376, 1393
 
 
\bibitem[Douglas et al.(2009)]{Douglas09} Douglas, L.~S.,  Bremer, M.~N., Stanway, E.~R.,  Lehnert, M.~D., Clowe, D.\  2009, \mnras, 400, 561
  
  \bibitem[Douglas et al.(2010)]{Douglas10} Douglas, L.~S., Bremer, 
M.~N., Lehnert, M.~D., Stanway, E.~R., 
\& Milvang-Jensen, B.\ 2010, \mnras, 409, 1155 
  
\bibitem[Finkelstein et al.(2009)]{Finkelstein09} Finkelstein, S.~L., Rhoads, J.~E., Malhotra, S., \& Grogin, N.\ 2009, \apj, 691, 465 


\bibitem[Gonzalez et al.(2011)]{Gonzalez11} Gonzalez, J.~E., 
Lacey, C.~G., Baugh, C.~M., Frenk, C.~S., 
\& Benson, A.~J.\ 2011, arXiv:1105.3731 


\bibitem[Hatsukade et al.(2011)]{Hatsukade11} Hatsukade, B., et 
al.\ 2011, \mnras, 411, 102

\bibitem[Hunt et al.(2005)]{Hunt05} Hunt, L., Bianchi, S., \& Maiolino, R.\ 2005, \aap, 434, 849 


\bibitem[Kennicutt et al.(1998)]{Kennicutt98} Kennicutt, R.~C., 
Jr., Stetson, P.~B., Saha, A., et al.\ 1998, \apj, 498, 181 


\bibitem[Knudsen et al.(2010)]{Knudsen10} Knudsen, K.~K., Kneib, 
J.-P., Richard, J., Petitpas, G., \& Egami, E.\ 2010, \apj, 709, 210 

\bibitem[\protect\citeauthoryear{Kohno}{2011}]{Kohno11} Kohno
K., 2011, EAS, 52, 23


\bibitem[Kreysa et al.(1998)]{Kreysa98} Kreysa, E., et al.\ 
1998, \procspie, 3357, 319 



\bibitem[Kruegel \& Siebenmorgen(1994)]{Kruegel94} Kruegel, E., \& Siebenmorgen, R.\ 1994, \aap, 288, 929 

\bibitem[Law et al.(2007)]{Law07} Law, D.~R., Steidel, C.~C., 
Erb, D.~K., et al.\ 2007, \apj, 656, 1 


\bibitem[Maiolino et al.(2004)]{Maiolino04} Maiolino, R., 
Schneider, R., Oliva, E., Bianchi, S., Ferrara, A., Mannucci, F., Pedani, 
M., \& Roca Sogorb, M.\ 2004, \nat, 431, 533 

\bibitem[McLure et al.(2009)]{McLure10} McLure, R.~J., 
Cirasuolo, M., Dunlop, J.~S., Foucaud, S., 
\& Almaini, O.\ 2009, \mnras, 395, 2196 




\bibitem[Mo \& White(2002)]{Mo02} Mo, H.~J., \& White, S.~D.~M.\ 2002, \mnras, 336, 112

\bibitem[Negrello et al.(2010)]{Negrello10} Negrello, M., et al.\ 
2010, Science, 330, 800

  
  \bibitem[Nilsson et al.(2007)]{Nilsson07} Nilsson, K.~K., et al.\ 2007, \aap, 471, 71


\bibitem[{{Oke} \& {Gunn}(1983)}]{Oke83}
{Oke}, J.~B. \& {Gunn}, J.~E. 1983, \apj, 266, 713

\bibitem[Overzier et al.(2009)]{Overzier08} Overzier, R.~A., Guo, 
Q., Kauffmann, G., De Lucia, G., Bouwens, R., 
\& Lemson, G.\ 2009, \mnras, 394, 577 


\bibitem[Priddey \& McMahon(2001)]{Priddey01} Priddey, R.~S., \& McMahon, R.~G.\ 2001, \mnras, 324, L17 



\bibitem[Riechers et al.(2010)]{Riechers10} Riechers, D.~A., et 
al.\ 2010, \apjl, 720, L131 

\bibitem[Rieke et al.(2009)]{Rieke09} Rieke, G.~H., 
Alonso-Herrero, A., Weiner, B.~J., et al.\ 2009, \apj, 692, 556 

\bibitem[Rigopoulou et al.(2010)]{Rigopoulou10} Rigopoulou, D., 
Magdis, G., Ivison, R.~J., et al.\ 2010, \mnras, 409, L7 

\bibitem[Schinnerer et al.(2008)]{Schinnerer08} Schinnerer, E., 
Carilli, C.~L., Capak, P., et al.\ 2008, \apjl, 689, L5 

\bibitem[Smail et al.(2002)]{Smail02} Smail I., Ivison R.J., Blain A.W., Kneib J.-P. \mnras, 331, 495 

\bibitem[\protect\citeauthoryear{{Solomon} \& {Vanden Bout}}{{Solomon} \&
  {Vanden Bout}}{2005}]{Solomon05}
{Solomon} P.~M.,  {Vanden Bout} P.~A.,  2005, ARA\&A, 43, 677


\bibitem[Sobral et al.(2012)]{Sobral12} Sobral, D., Smail, I., 
Best, P.~N., et al.\ 2012, arXiv:1202.3436 


\bibitem[Stanway et al.(2008)]{Stanway08a} Stanway, E.~R., Bremer, 
M.~N., \& Lehnert, M.~D.\ 2008, \mnras, 385, 493 

\bibitem[\protect\citeauthoryear{{Stanway}, {Bremer}, {Davies}, {Birkinshaw},
  {Douglas} \& {Lehnert}}{{Stanway} et~al.}{2008}]{Stanway08}
{Stanway} E.~R.,  {Bremer} M.~N.,  {Davies} L.~J.~M.,  {Birkinshaw} M.,
  {Douglas} L.~S.,    {Lehnert} M.~D.,  2008, \apjl, 687, L1


\bibitem[Stanway et al.(2010)]{Stanway10} Stanway, E.~R., Bremer, 
M.~N., Davies, L.~J.~M., \& Lehnert, M.~D.\ 2010, \mnras, 407, L94 

\bibitem[\protect\citeauthoryear{{Stark}, {Ellis}, {Bunker}, {Bundy},
  {Targett}, {Benson} \& {Lacy}}{{Stark} et~al.}{2009}]{Stark09}
{Stark} D.~P.,  {Ellis} R.~S.,  {Bunker} A.,  {Bundy} K.,  {Targett} T.,
  {Benson} A.,    {Lacy} M.,  2009, \apj, 697, 1493

\bibitem[Thomas et al.(2010)]{Thomas10} Thomas, D., Maraston, 
C., Schawinski, K., Sarzi, M., \& Silk, J.\ 2010, \mnras, 404, 1775 

\bibitem[\protect\citeauthoryear{{Vanzella}, {Giavalisco}, {Dickinson},
  {Cristiani}, {Nonino}, {Kuntschner}, {Popesso}, {Rosati}, {Renzini}, {Stern},
  {Cesarsky}, {Ferguson} \& {Fosbury}}{{Vanzella} et~al.}{2009}]{vanzella09}
{Vanzella} E.,  {Giavalisco} M.,  {Dickinson} M.,  {Cristiani} S.,  {Nonino}
  M.,  {Kuntschner} H.,  {Popesso} P.,  {Rosati} P.,  {Renzini} A.,  {Stern}
  D.,  {Cesarsky} C.,  {Ferguson} H.~C.,    {Fosbury} R.~A.~E.,  2009, \apj,
  695, 1163

\bibitem[\protect\citeauthoryear{{Verma}, {Lehnert}, {F{\"o}rster Schreiber},
  {Bremer} \& {Douglas}}{{Verma} et~al.}{2007}]{Verma07}
{Verma} A.,  {Lehnert} M.~D.,  {F{\"o}rster Schreiber} N.~M.,  {Bremer} M.~N.,
    {Douglas} L.,  2007, \mnras, 377, 1024

\bibitem[Wang et al.(2008)]{Wang08} Wang, R., et al.\ 2008, 
\apj, 687, 848

\bibitem[Webb et al.(2003)]{Webb03} Webb, T.~M., et al.\ 2003, \apj, 582, 6 

\bibitem[Wei{\ss} et al.(2007)]{Weib07} Wei{\ss}, A., Downes, D., Neri, R., Walter, F., Henkel, C., Wilner, D.~J., Wagg, J., \& Wiklind, T.\ 2007, \aap, 467, 955 




\end{thebibliography}
\end{document}